\documentclass[aps,prd,twocolumn]{revtex4-2}
\usepackage{hyperref}
\usepackage{amsmath}
\usepackage{amssymb}
\usepackage{graphicx}
\usepackage{xcolor}

\begin{document}

\title{False vacuum decay in triamond lattice gauge theory}

\author{Ali H. Z. Kavaki}\email{alihzk@yorku.ca}
\author{Randy Lewis}\email{randy.lewis@yorku.ca}
\affiliation{Department of Physics and Astronomy, York University, Toronto, Ontario, Canada, M3J 1P3}

\date{July 1, 2025}

\begin{abstract}
The transition from a false vacuum to the true vacuum is a real-time phenomenon of interest in many contexts.
It represents a special challenge for strongly interacting non-Abelian gauge theories because standard
spacetime lattices incorporate imaginary time from the outset.
To attain real-time phenomena, Hamiltonian lattice methods are being developed for quantum computers.
The present work considers SU(2) gauge theory on a minimal lattice in three spatial dimensions, where round-the-world
strings called torelons can travel and interact.
This minimal 3D lattice has a triamond structure whose properties are elucidated by calculating the spectrum of torelon states.
Then, by introducing a twisted boundary condition, real-time evolution is used to demonstrate the decay of a false vacuum.
Calculations in the present work are done on classical computers except for one benchmark study of imaginary time evolution
that ran on the {\scshape ibm\_brisbane} quantum computer.
\end{abstract}

\maketitle

\section{Introduction}

The classic picture of false vacuum decay is a double-well potential energy function where the shallower well
(called the false vacuum) is separated from the deeper well (called the true vacuum) by an energy barrier
\cite{Coleman:1977py,Callan:1977pt}.
A system that begins in the false vacuum can later be found in the true vacuum with some probability.
For the corresponding scenario in a quantum field theory,
the decay rate can be calculated by using a semiclassical approximation and imaginary time,
but that approach does not explain the physical mechanism of the decay.

In contrast, real-time simulations would allow direct observation of the vacuum decay mechanism,
perhaps showing bubbles of true vacuum appearing and nucleating to displace the false vacuum.
Various approaches are being explored by several authors
\cite{Braden:2018tky,Ai:2019fri,Mou:2019gyl,Zenesini:2023afv,Nishimura:2023dky,Batini:2023zpi},
including quantum simulations of Ising models
\cite{Sinha:2021aua,Lagnese:2021grb,Lagnese:2023xjg,Vodeb:2024tvo,Luo:2025qlg} and the Schwinger model \cite{Zhu:2024dvz}.
In the present work, we take a first step toward using a non-Abelian lattice gauge theory for the real-time simulation of false vacuum decay.
We use a pure gauge theory; no matter fields are required.

Lattice gauge theory is a direct computational method for calculations in non-Abelian theories such as quantum chromodynamics (QCD).
Computations of the masses, decay constants and form factors of QCD hadrons have become precision results of the utmost
importance to the particle and nuclear physics communities
\cite{FlavourLatticeAveragingGroupFLAG:2024oxs,ParticleDataGroup:2022pth}.
The starting point for standard computations is a four-dimensional Euclidean spacetime lattice that provides convenient access to
imaginary time calculations but no access to real-time simulations.

A promising approach to real-time physics is to rewrite the theory as a Hamiltonian problem, keeping time as a continuous
variable and discretizing only the spatial components onto a three-dimensional lattice \cite{Kogut:1974ag}.
The length of the state vector grows exponentially with the lattice size but, in principle, it can be stored efficiently in the
qubit register of a quantum computer \cite{Davoudi:2020yln}.
Unfortunately, the non-Abelian version of Gauss's law must now be imposed as a constraint that is external to the Hamiltonian,
and finding the best approach to this issue is a high-priority topic for several research groups
\cite{Raychowdhury:2018osk,Stryker:2018efp,Emonts:2018puo,Kaplan:2018vnj,Davoudi:2022xmb,Kane:2022ejm,DAndrea:2023qnr,Carena:2024dzu,Ball:2024xmw,Kadam:2024zkj,Fontana:2024rux,Grabowska:2024emw,Burbano:2024uvn,Ballini:2024qmr}.
Nevertheless, non-Abelian theories have already been put successfully onto quantum computers
\cite{Klco:2019evd,ARahman:2021ktn,ARahman:2022tkr,Ciavarella:2021nmj,Ciavarella:2021lel,Alam:2021uuq,Illa:2022jqb,Gustafson:2022xdt,Atas:2021ext,Farrell:2022wyt,Atas:2022dqm,Farrell:2022vyh,Ciavarella:2023mfc,Kavaki:2024ijd,Ciavarella:2024fzw,Turro:2024pxu,Hayata:2024fnh,Ciavarella:2024lsp,Than:2024zaj},
though the lattices are necessarily small and typically have just one or two spatial dimensions.

In three dimensions, each site on a cubic lattice is touched by six gauge links, pointed in directions $\pm\hat i$, $\pm\hat j$, $\pm\hat k$ along the standard Cartesian axes.
There are multiple ways that six specified SU(2) representations can be combined to satisfy Gauss's law.
This means additional qubits are required at each site to define the partial sums needed for a complete definition of the quantum state.
The triamond lattice is an alternative that spans three-dimensional space and yet has only three gauge links touching each site
\cite{Kavaki:2024ijd}.
This means the triamond gauge fields are fully defined by the gauge link qubits with no need for extra qubits at the sites.
The triamond lattice is strongly isotropic \cite{Sunada,Suizu,Kavaki:2024ijd}
and provides an efficient way to build a three-dimensional lattice gauge theory.
A similar approach in two dimensions leads to hexagonal lattices \cite{Muller:2023nnk,Ebner:2023ixq,Ebner:2024qtu}.

\begin{figure}
\includegraphics[width=85mm]{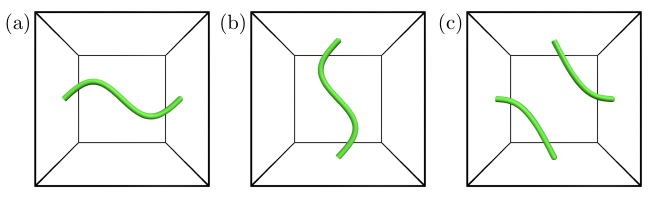}
\caption{Closed path of round-the-world color flux, called a torelon, in a cubic spatial volume with periodic boundary conditions.
         (a) One horizontal torelon. (b) One vertical torelon. (c) One diagonal torelon.
}\label{fig:torelons}
\end{figure}

In the absence of matter fields, gauge invariance requires any flow of non-Abelian charges (such as the colors of QCD) to form
closed paths on the lattice.
Localized particles called glueballs are gauge-invariant objects described by a superposition of localized paths.
If the lattice has periodic boundary conditions, there can also be paths that go all the way around the lattice to close on
themselves from behind as shown in Fig.~\ref{fig:torelons}, and these are called torelons
\cite{Michael:1986cj,Michael:1989mf,Michael:1989vh,Juge:2003vw,Meyer:2004vr,GarciaPerez:2013idu,GarciaPerez:2018fkj,Athenodorou:2021vkw}.
A torelon is a closed path of color flux that is not homotopic to a point.
Torelons become nondynamical heavy states as the lattice volume is taken to infinity, but on small lattices they are
central to the physics.

In SU(2) gauge theory with periodic boundary conditions, the presence or absence of a torelon in the $\hat i$ direction is a good
quantum number, and the same is true for the $\hat j$ direction and the $\hat k$ direction.
A pair of torelons along the same direction can annihilate each other, but a single torelon cannot disappear.
Torelons can move across the lattice and interact with each other through the SU(2) gauge fields.
A pair of torelons along two different directions cannot annihilate each other but they can combine to form a bound state
that we will call a diagonal torelon,
corresponding to a single path winding around the lattice in both of the original directions before returning to itself,
as shown in Fig.~\ref{fig:torelons}(c).

For a universe containing just one torelon (suppose it is a diagonal torelon in the $\hat i$ and $\hat k$ directions)
the ground state has no objects other than the stationary torelon.
That diagonal torelon has the same quantum numbers as a pair of separated
torelons, one in each of the $\hat i$ and $\hat k$ directions, but this
two-torelon state has higher energy.
If the universe is periodic in all three directions, then the ground state is stable.
However, a twisted boundary condition in the $\hat j$ direction means $\hat i$ and $\hat k$ become interchanged for an object that travels
completely around the $\hat j$ direction.  This means the diagonal torelon state (false vacuum) can decay to a no-torelon state (true vacuum)
and that these two states are separated by the two-torelon state (energy barrier).
This is the false vacuum decay process that will be investigated in the present work.

Section~\ref{sec:triamond} defines SU(2) gauge theory on a triamond lattice, explaining the specific geometry that will
be used for the current study and how it can be mapped to a ladder of square paths.
Section~\ref{sec:spectrum} examines the spectrum of torelon states on a triamond lattice of three unit cells, discussing
the subtle relationship between energy and momentum.
Section~\ref{sec:qite} presents our method for running quantum imaginary time evolution on an IBM quantum computer for a lattice
with 12 square plaquettes, shows the results of those quantum computations, and contemplates the extension to a triamond lattice.
Section~\ref{sec:realtime} contains our use of real-time evolution to study false vacuum decay, which is built on the fact that
the triamond lattice is truly three-dimensional.
Section~\ref{sec:outlook} provides an outlook toward possible next steps.

\section{The triamond lattice}\label{sec:triamond}

A primary motivation for choosing a triamond lattice rather than a cubic lattice is to reduce the number of
gauge links touching each site from 6 to 3 \cite{Kavaki:2024ijd}.
Having only three gauge links at a site means the combined SU(2) value of any pair of links is determined uniquely by the value of the
third gauge link.
In contrast, the combined SU(2) value of two links on a cubic lattice is not unique because the other four links can
be combined in many ways, and this means extra qubits would be required at each lattice site to fully specify the quantum state.

\begin{figure}
\includegraphics[width=85mm]{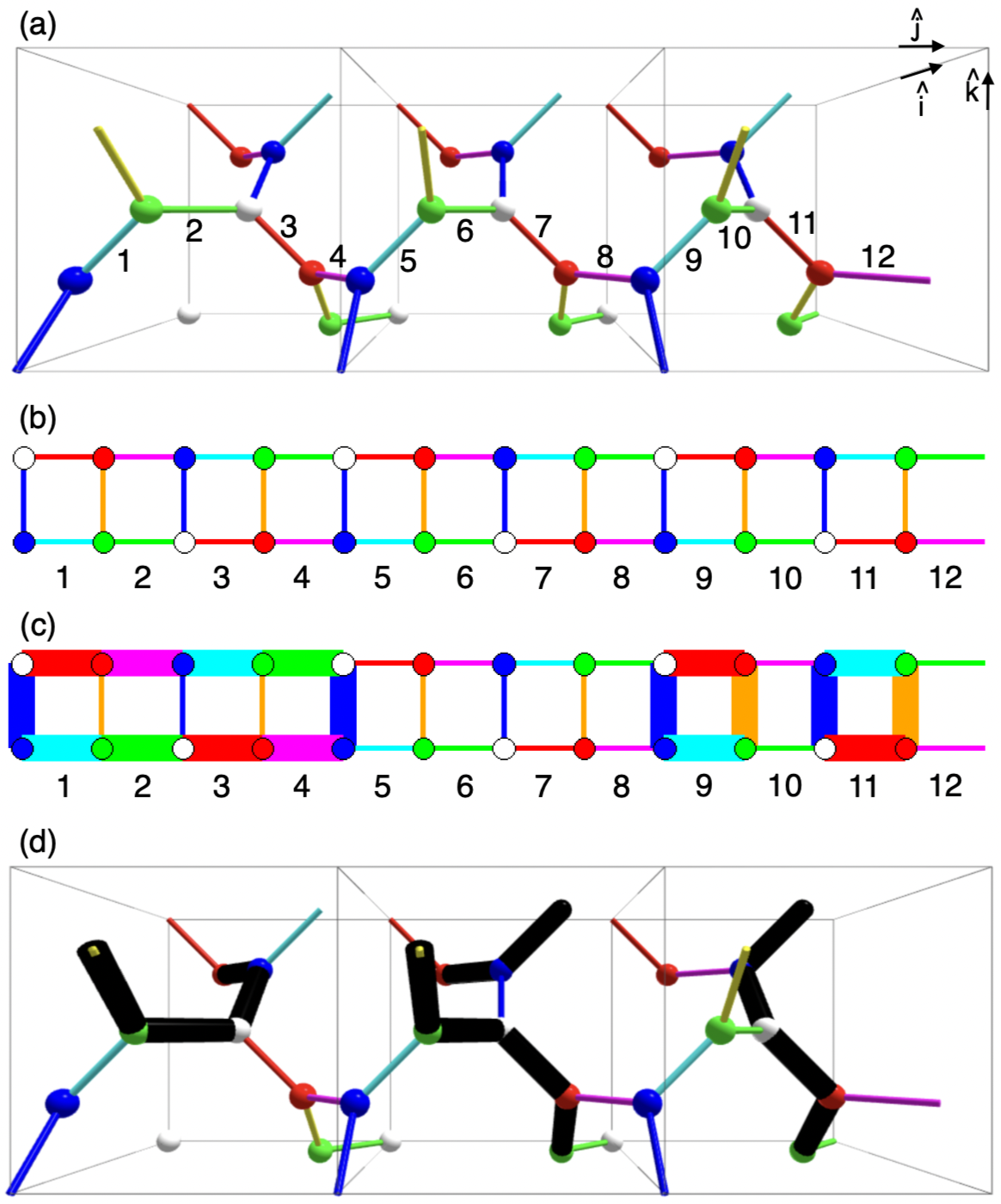}
\caption{(a) Triamond lattice comprising three unit cells with periodic boundary conditions in all three directions.
Twelve of the 36 gauge links are numbered so they can be referenced in the text.
(b) A flattened map of the same triamond lattice with the same 12 labels.
(c) The same flattened map with two plaquettes shown in thick lines, namely a 10-sided plaquette at 1,2,3,4 and an 8-sided
plaquette at 9,11.
(d) Thick black paths show a horizontal, diagonal and vertical torelon in the left, center and right unit cells, respectively.
}\label{fig:LatticeLayout}
\end{figure}

The triamond lattice used in this work is shown in Fig.~\ref{fig:LatticeLayout}(a).
To view a triamond lattice from many different vantage points, please see our two-minute video \cite{triamondvimeo}.
At each lattice site, the three gauge links lie in a plane and are equally spaced.
The plane at a lattice site is always orthogonal to the line connecting opposite corners of a unit cell, which means there
are four different planes in a triamond lattice.
The lattice sites in Fig.~\ref{fig:LatticeLayout} are shown in white, red, green and blue to label the plane at each site.

In a cubic lattice all gauge links are on the edges of a unit cell, but the triamond lattice is not so sparse and
has gauge links inside each unit cell.
Specifically, every gauge link lies along one of the following six directions,
\begin{align}
\hat r &= \frac{\hat j - \hat k}{\sqrt{2}}, &
\hat g &= \frac{\hat i - \hat j}{\sqrt{2}}, &
\hat b &= \frac{\hat k - \hat i}{\sqrt{2}}, \nonumber
\end{align}
\begin{align}
\hat c &= \frac{\hat j + \hat k}{\sqrt{2}}, &
\hat m &= \frac{\hat i + \hat j}{\sqrt{2}}, &
\hat y &= \frac{\hat k + \hat i}{\sqrt{2}}, \label{eq:colorvectors}
\end{align}
corresponding respectively to the red, green, blue, cyan, magenta and yellow gauge links in Fig.~\ref{fig:LatticeLayout}
with $\hat i$, $\hat j$ and $\hat k$ being the standard orthonormal unit vectors.

Because of the periodic boundaries, all connections between gauge links in Fig.~\ref{fig:LatticeLayout}(a) can be
shown in a flat map as displayed in Fig.~\ref{fig:LatticeLayout}(b).
The map has the form of a 12-rung ladder that is periodic only in the horizontal direction.
For example, the links along one side of the ladder (numbered from 1 to 12 in Fig.~\ref{fig:LatticeLayout}) lie on a
corkscrew trajectory around the periodic triamond lattice.

On a large triamond lattice the smallest closed paths are 10-sided, and several examples can be found in
Fig.~\ref{fig:LatticeLayout}(a).
They are called elementary plaquettes.
Every elementary plaquette omits one of the gauge link directions from Eq.~(\ref{eq:colorvectors})
but includes two links in each of the other five directions.
Any triamond lattice comprising $N$ unit cells has $12N$ gauge links and $12N$ elementary plaquettes.

On our small triamond lattice, some of the elementary plaquettes wrap around a periodic boundary and affect the same link twice.
In the special case where the twice-affected link retains its original value, the result is pair creation of two torelons
that we will refer to as an 8-sided plaquette.
Figure~\ref{fig:LatticeLayout}(c) shows examples of 10-sided and 8-sided plaquettes.

The Hamiltonian of SU(2) gauge theory is a sum of a color-electric term and a color-magnetic term \cite{Kavaki:2024ijd},
\begin{equation}
H_{\text{triamond}} = H_E + H_B,
\end{equation}
with
\begin{eqnarray}
H_E &=& \frac{8\sqrt{2}a^3g^2}{3}\sum_{n=\text{links}}\text{Tr}\left(E_x^2(n)+E_y^2(n)+E_z^2(n)\right),~~~~ \\
H_B &=& -\frac{2\sqrt{2}}{g^2a}\sum_{\vec w=\text{white}}\sum_{s=1}^6{\cal P}_s(\vec w),
\end{eqnarray}
where $g$ is the SU(2) coupling and $a$ is the lattice spacing, i.e.\ the length of each gauge link.
The sum in the color-electric term runs over all gauge links on the lattice, with $\vec E$ representing the color-electric field.
The double sum in the color-magnetic term runs over all plaquettes on the lattice, organized here as six plaquettes per white site
as described in Ref.~\cite{Kavaki:2024ijd}.
In particular, ${\cal P}_s(\vec w)$ is the trace of the product of gauge links around the $s$'th plaquette at white site $\vec w$.

Any eigenstate of the color-electric Hamiltonian is fully defined by the SU(2) quantum numbers of the individual gauge links,
$j_1$, $j_2$, $j_3$, \ldots, $j_{12N}$.
In terms of these states, the diagonal matrix elements of the Hamiltonian are
\begin{equation}
\left<\{j\}\right|H_E\left|\{j\}\right> = \frac{8\sqrt{2}g^2}{3a}\sum_{n=1}^{12N}j_n(j_n+1)
\end{equation}
and the off-diagonal matrix elements are obtained from a product over all lattice sites on the perimeter of a plaquette,
\begin{eqnarray}
\left<\{J\}\right|{\cal P}_s(\vec w)\left|\{j\}\right> &=& \prod_{\text{perimeter}}(-1)^{j_e+j_f+J_b}\sqrt{2J_f+1} \nonumber \\
&& \sqrt{2j_f+1}\left\{\begin{array}{ccc} j_e & j_f & j_b \\ \frac{1}{2} & J_b & J_f \end{array}\right\},
\end{eqnarray}
which matches the definition in Ref.~\cite{Muller:2023nnk}
and differs from Refs.~\cite{Zache:2023dko,Hayata:2023puo,Kavaki:2024ijd} by a factor of $(-1)^{1/2}$ inside the product.

The continuous SU(2) symmetry means each gauge link has an infinite basis of options, $j\in\{0,\frac{1}{2},1,\frac{3}{2},2,\ldots\}$.
Use of a finite number of qubits necessitates a truncation, and in the present work we retain only $j\in\{0,\frac{1}{2}\}$
which corresponds to one qubit per gauge link.
Gauss's law restricts the options at every lattice site and allows the complete state of the lattice in Fig.~\ref{fig:LatticeLayout}
to be determined with only 13 qubits.
A specific choice is the 12 numbered gauge links plus any single link from the other side of the ladder.
This Hilbert space has dimension $2^{13}$ and consists of two orthogonal sectors according to the presence or absence of a torelon
in the long $\hat j$ direction.
Because long torelons correspond to heavy states, we will focus on the sector without a $\hat j$ torelon.
This leaves us with a Hilbert space of dimension $2^{12}$ that is spanned by just the 12 numbered gauge links of
Fig.~\ref{fig:LatticeLayout}.

The mapping of a triamond lattice to a ladder of square paths is quite convenient.
The three-dimensional aspects of the triamond lattice appear through the 10-sided and 8-sided plaquettes that involve more than just
nearest-neighbor interactions among the squares.

\section{The torelon spectrum}\label{sec:spectrum}

\begin{figure}
\includegraphics[width=85mm]{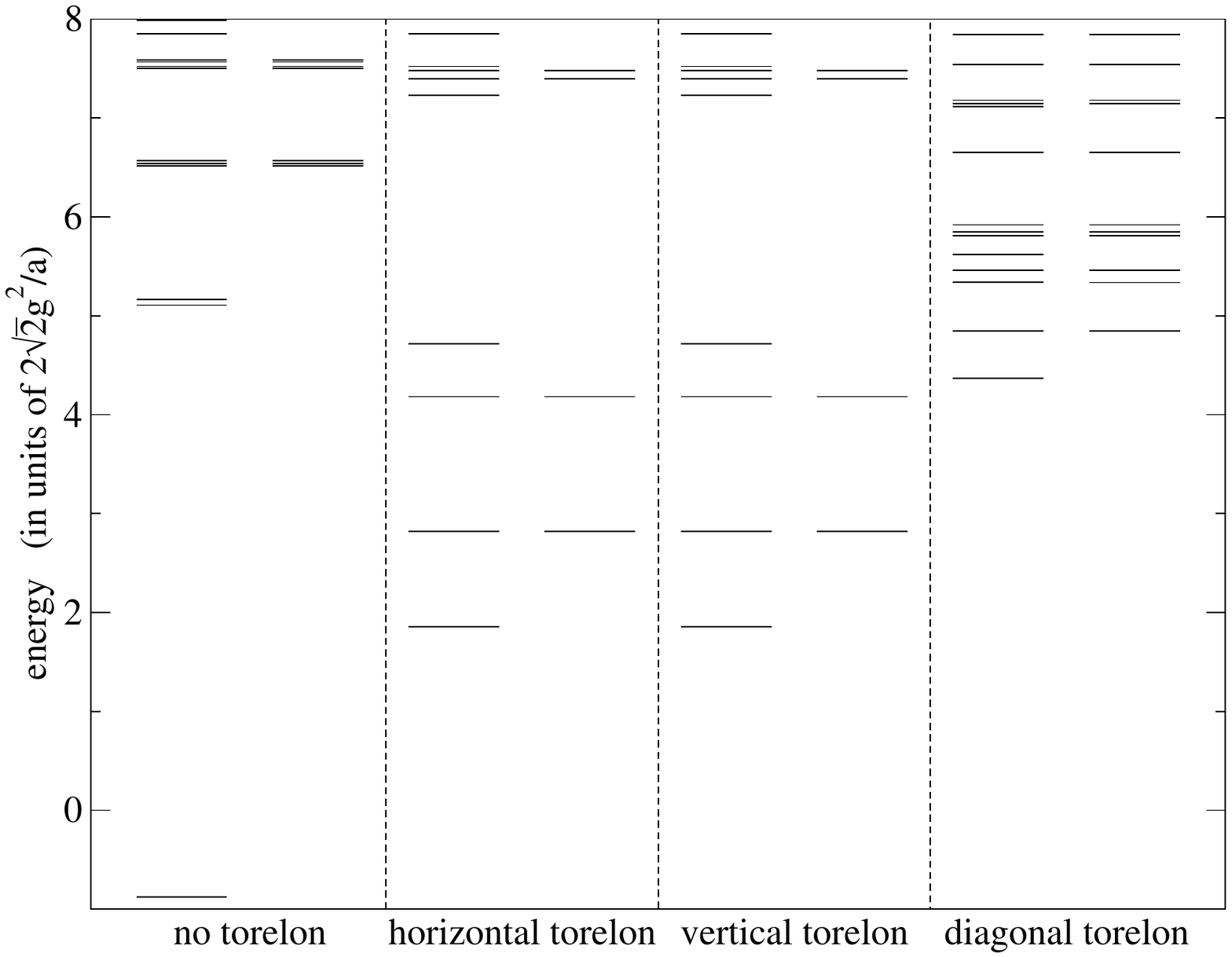}
\caption{Smallest eigenvalues of SU(2) gauge theory on a triamond lattice having three unit cells in a row and periodic
boundary conditions.  The gauge theory is truncated such that each gauge link has $j\in\{0,\frac{1}{2}\}$ and the gauge
coupling is $g=1.3$.
Numerical values from this graph are provided in Appendix~\ref{app:valsvecs}.
}\label{fig:spectrum}
\end{figure}

Construction of the explicit $4096\times4096$ Hamiltonian matrix for the triamond lattice having three unit cells allows
a direct determination of all eigenvalues and eigenvectors by standard classical computing methods.
The smallest eigenvalues are displayed in Fig.~\ref{fig:spectrum} for a specific choice of the gauge coupling.
Important insights can be gained by examining and interpreting the corresponding eigenstates.

To begin, notice that the vacuum state is significantly lighter than all other states, as expected.
In the strong-coupling limit the vacuum would have an energy of zero, but away from that limit it is pushed somewhat
below zero.
The vacuum state is dominated by the bare vacuum where every gauge link has $j=0$, and the leading corrections arise from
linear combinations of single-plaquette terms that are translationally invariant on the lattice.

The first excited state is degenerate and represents one zero-momentum torelon in either of the two short directions on the
lattice, $\hat i$ or $\hat k$.
There are six locations for a horizontal torelon, and the groupings into six momentum states are evident in Fig.~\ref{fig:spectrum}.
Degeneracies within the horizontal sector are due to the equal energies of forward and backward momenta.

All energies in the vertical torelon sector are identical to the horizontal case because of a lattice symmetry.
To be precise, it is a screw symmetry that rotates the lattice by $\pi/2$ around $\hat j$ while also pushing it forward in the
$\hat j$ direction by $\frac{1}{4}$ of a unit cell.
On the ladder map in Fig.~\ref{fig:LatticeLayout}, the screw symmetry is a shift by one square.

The lowest energy state of a diagonal torelon is translationally invariant and represents a diagonal torelon with zero momentum.
Relative to the vacuum, the diagonal torelon energy is less than the sum of a horizontal torelon and a vertical torelon, which is
expected for sufficiently large gauge coupling because a pair of basic torelons needs eight gauge links whereas a diagonal torelon needs only six.
This is a reminder that some of the excited states in the diagonal torelon sector of Fig.~\ref{fig:spectrum}
can have a large overlap with a pair of spatially separated torelons.

The 12 lowest energy states in the diagonal torelon sector tell an important story about how momentum is handled on a
triamond lattice.
These 12 lowest states are the energy eigenstates that correspond to placing a diagonal torelon at any of the 12 allowed
positions on the lattice.
In the ladder map of Fig.~\ref{fig:LatticeLayout}(b), they appear as superpositions of a single six-link rectangle at each of the 12 possible locations.
Notice that half of these rectangles are doubly blue and nonyellow while the other half are nonblue and doubly yellow.

The translation along the 12-step ladder of Fig.~\ref{fig:LatticeLayout}(b) involves a $\pi/2$ rotation with each step along the
$\hat j$ direction and is called a screw translation.
The Fourier transform of this screw translation results in eigenstates of screw momentum that are also eigenstates of energy,
and these are precisely the states shown in Fig.~\ref{fig:spectrum}.
Importantly, the 12 values of screw momentum contain six values of linear momentum for the nonyellow torelons and six
values for the nonblue torelons.
On the original triamond lattice, the nonblue torelon lies along a lattice diagonal that is spatially
orthogonal to the nonyellow torelon.  A nonblue torelon path is sketched in Fig.~\ref{fig:LatticeLayout}(d).

Excited states in the no-torelon sector include the momentum states of various operators.
To see detailed numerical results for energy eigenvalues and eigenvectors from all sectors of Fig.~\ref{fig:spectrum},
please consult App.~\ref{app:valsvecs}.

\section{Imaginary time evolution}\label{sec:qite}

Real-time evolution is a central goal for the quantum computation of gauge theories, but it requires
creation of an appropriate initial state.
Imaginary time evolution is a valuable method for creating the initial state.
Here we implement a successful determination of the ground state on 12 qubits of the {\scshape ibm\_brisbane}
quantum computer for a simplified SU(2) Hamiltonian.
This success provides insight into the practical challenge of coding a triamond lattice into a quantum computer
that has limited connectivity among its qubits.

Although the triamond lattice for three unit cells can be mapped onto the ladder of Fig.~\ref{fig:LatticeLayout}(b),
the three-dimensional nature of the triamond structure remains evident through the large
plaquettes shown in Fig.~\ref{fig:LatticeLayout}(c).
A simpler system is obtained by using the ladder directly as a quasi-one-dimensional lattice having only square
plaquettes and no connection to the three-dimensional triamond structure.
This ladder of simple square plaquettes has been used in several quantum computations for SU($N$) gauge theory
\cite{Klco:2019evd,ARahman:2021ktn,ARahman:2022tkr,Ciavarella:2021nmj,Ciavarella:2021lel}.
Imaginary time evolution \cite{Motta:2019yya} for a nonperiodic ladder of only three rungs was run on a
quantum computer recently \cite{Kavaki:2024ijd}.
Here we will study the 12-rung periodic ladder.

\begin{figure}
\includegraphics[width=85mm]{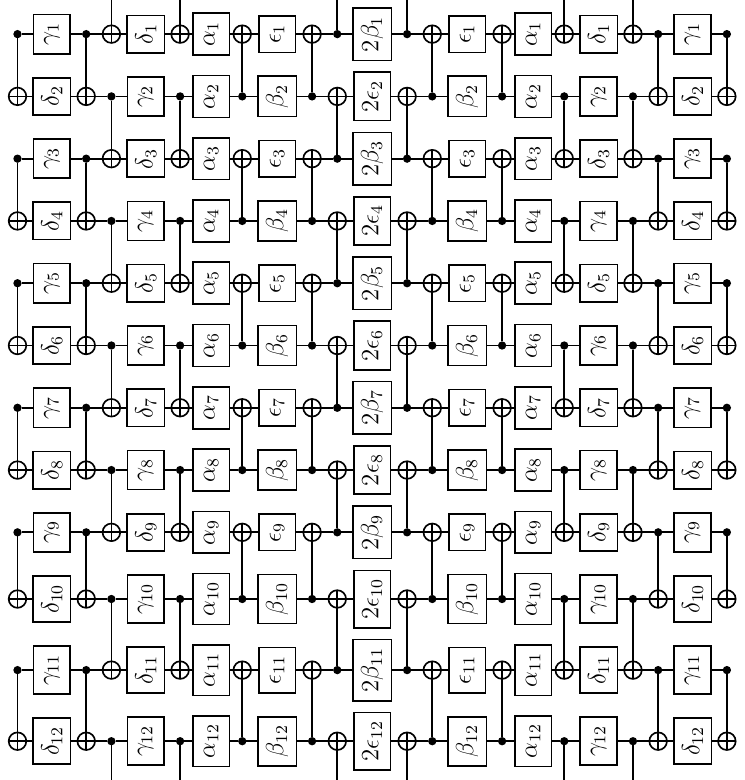}
\caption{One step of imaginary time evolution for the Hamiltonian of Eq.~(\ref{eq:Hsquare}) and the ansatz of
         Eq.~(\ref{eq:ansatz}).
         All boxes represent an RY$(\theta)$ gate with the angle $\theta$ given in the box.
         The circuit is for a closed loop of 12 qubits, so CNOT gates emerging from the top of the diagram are continued at
         the bottom.
         Multiple steps of this circuit ran on a periodic ring of 12 qubits on {\scshape ibm\_brisbane}.
}\label{fig:circuit}
\end{figure}

In units of $2/g^2$, the SU(2) Hamiltonian for square plaquettes with gauge coupling $x=2/g^4$ is
\begin{eqnarray}
H_\square &=& \frac{27}{2} - \frac{3}{8}\sum_{j=1}^{12}(2Z_j+Z_jZ_{j+1}) \nonumber \\
  & & - \frac{x}{8}\sum_{j=1}^{12}(9+3Z_{j-1}+3Z_{j+1}+Z_{j-1}Z_{j+1})X_j, \hspace{5mm} \label{eq:Hsquare}
\end{eqnarray}
where $X_j$ and $Z_j$ are Pauli gates acting on the $j$th qubit.
Imaginary time evolution from any initial state is
\begin{equation}
\left|\Psi(\tau)\right> = e^{-\tau H_\square}\left|\Psi(0)\right> = r^\prime e^{-i\tau A}\left|\psi\right>,
\end{equation}
where the normalizing factor can be obtained from
\begin{equation}
r^\prime = r(1-\tau\left<\psi\right|H_\square\left|\psi\right>) + O(\tau^2).
\end{equation}
Because the Hamiltonian is purely real, the matrix $A$ must be purely imaginary.
This means each term has an odd number of $Y$ gates.
For a two-qubit Hamiltonian, the most general expression is
\begin{eqnarray}
A &=&   a_{iy}Y_1 + a_{xy}X_2Y_1 + a_{zy}Z_2Y_1 \nonumber \\
  & & + a_{yi}Y_2 + a_{yx}Y_2X_1 + a_{yz}Y_2Z_1.
\end{eqnarray}
For the 12-qubit Hamiltonian of Eq.~(\ref{eq:Hsquare}),
the most general expression has many terms but we might expect
only a subset of them to dominate the physics.
In particular, since $H_\square$ has only single-qubit terms, adjacent-pair terms, and adjacent-triple terms, we
anticipate that $A$ will be predominantly local.
Specifically, the single-qubit and adjacent-pair terms are expected to be most important, with adjacent-triple terms being less important.
Since adjacent-triple terms would require swap gates when implemented on IBM's heavy-hex hardware architecture, we propose to neglect
these subleading terms.  Therefore
our ansatz has only single-qubit and adjacent-pair terms,
\begin{eqnarray}
A &=& \sum_{j=1}^{12}\bigg( (a_y)_jY_j + (a_{xy})_jX_{j+1}Y_j + (a_{zy})_jZ_{j+1}Y_j \nonumber \\
  & & + (a_{yx})_jY_jX_{j-1} + (a_{yz})_jY_jZ_{j-1} \bigg). \label{eq:ansatz}
\end{eqnarray}
Notice that each qubit $j$ gets its own set of coefficients $(a_\bullet)_j$, thus allowing our error mitigation to make
no assumption about individual physical qubits having similar noise profiles.
The coefficients $(a_\bullet)_j$ are determined from state tomography as explained in detail in Appendix~\ref{app:qite}.
The exponentiation of each term in $A$ is accomplished by using
\begin{eqnarray}
e^{-i\theta Y_j} &=& RY_j(2\theta), \\
e^{-i\theta X_jY_k} &=& CX_{kj}RY_k(2\theta)CX_{kj}, \\
e^{-i\theta Z_jY_k} &=& CX_{jk}RY_k(2\theta)CX_{jk}.
\end{eqnarray}
The complete time evolution $e^{-i\tau A}\left|\psi\right>$ can be constructed as a second-order Trotter circuit
with terms ordered in a way that minimizes the number of entangling gates, as shown in Fig.~\ref{fig:circuit}.
Notice that CNOT gates at the end of a time step (right edge of the figure) will cancel with CNOT gates at the
beginning of the next time step (left edge of the figure).
The rotation angles in the circuit are
\begin{eqnarray}
\alpha_j   &=& (a_y)_j\Delta\tau, \\
\beta_j    &=& (a_{yx})_j\Delta\tau, \\
\gamma_j   &=& (a_{xy})_j\Delta\tau, \\
\delta_j   &=& (a_{yz})_j\Delta\tau, \\
\epsilon_j &=& (a_{zy})_j\Delta\tau,
\end{eqnarray}
and their numerical values are different at each time step in the quantum circuit.
Values from previous time steps are stored in a list for reuse, and values for the new time step are
computed from the previous time steps according to Appendix~\ref{app:qite}.

\begin{figure}
\includegraphics[width=85mm]{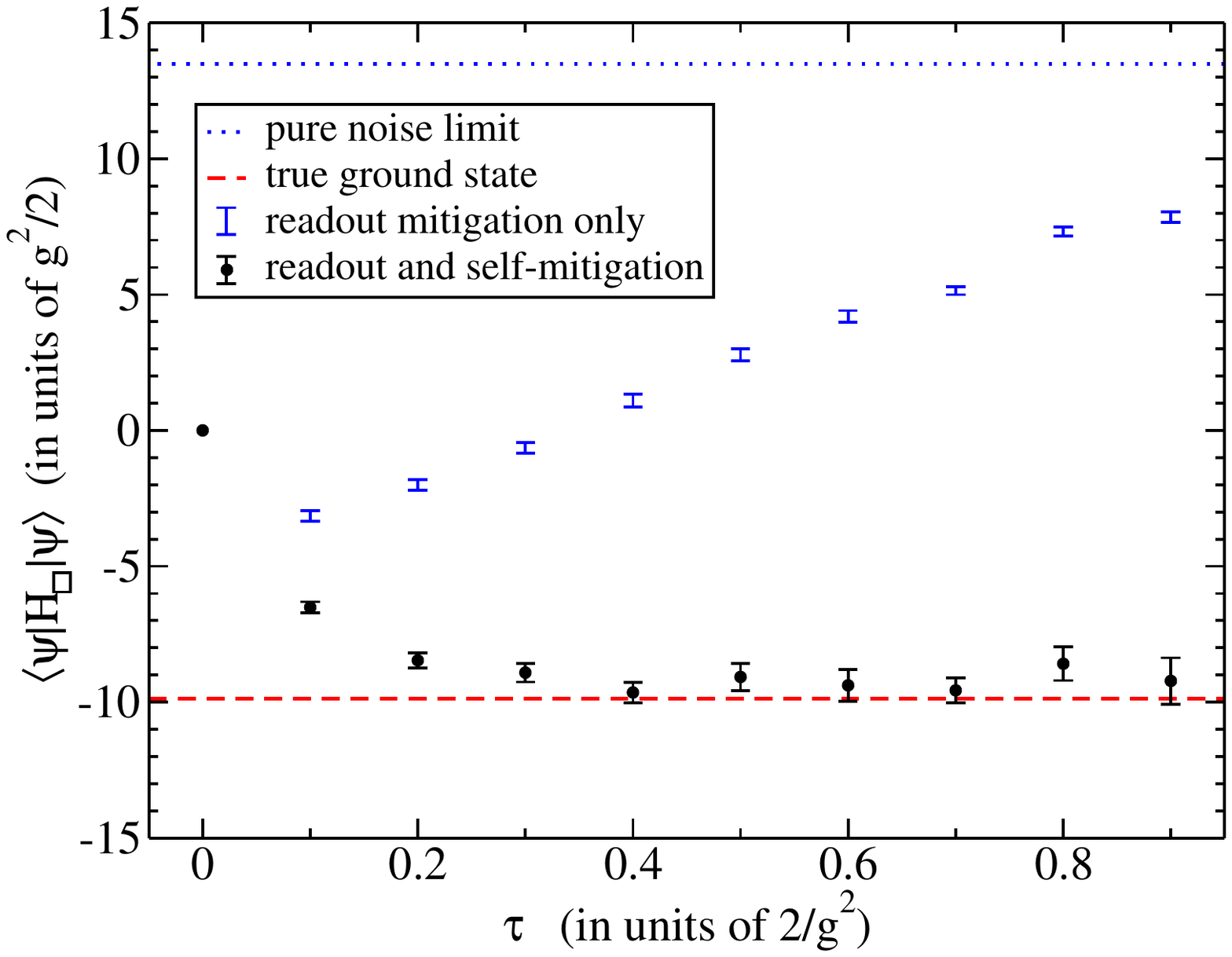}
\caption{Imaginary time evolution of SU(2) gauge theory on a periodic 12-plaquette ladder.
The Hamiltonian is Eq.~(\ref{eq:Hsquare}) with gauge coupling $g=1$.
The true ground state is obtained successfully because of the use of self-mitigation.
The code ran on 12 qubits of {\scshape ibm\_brisbane}.
Numerical values from this graph are provided in Table~\ref{tab:brisbane}.
}\label{fig:plot12mitC}
\end{figure}

The heavy-hex layout of {\scshape ibm\_brisbane} provides precisely the loop of 12 qubits that is required for
this computation.  Because of noisy hardware, a direct computation with mitigation of readout errors, which means mitigation
for the final measurement of each qubit, was unable to obtain the true ground state, as shown in Fig.~\ref{fig:plot12mitC}.
Self-mitigation was found to overcome this problem for two-qubit computations in Ref.~\cite{Kavaki:2024ijd},
and Fig.~\ref{fig:plot12mitC} shows that it is equally successful in the 12-qubit case.

The basic idea of self-mitigation \cite{ARahman:2022tkr} is to create a mitigation circuit that is very similar to the physics circuit.
The true result of the mitigation circuit is known in advance, so running that circuit determines the errors being
made by the hardware.  The measured errors are then used to rescale the original physics circuit of interest.
Specifically, if the physics circuit has $N$ steps forward in time, then the mitigation circuit has $N/2$ steps
forward followed by $N/2$ steps backward, thus arriving at the initial state modulo hardware errors.
An odd number for $N$ presents no problem because we are using second-order Trotter steps, which means each step
is already symmetric as seen in Fig.~\ref{fig:circuit} and can therefore readily be made half forward and half backward.

Details of the self-mitigation method can be found in Refs.~\cite{ARahman:2022tkr,Kavaki:2024ijd}.
Other uses and extensions of this approach to mitigation can be found in Refs.~\cite{Farrell:2022wyt,Atas:2022dqm,Ciavarella:2023mfc,Farrell:2024fit,Charles:2023zbl,Farrell:2023fgd,Asaduzzaman:2023wtd,Hidalgo:2023wzr,Kiss:2024foh,Ciavarella:2024fzw,Koenig:2024bom,Gustafson:2024jop,Turro:2024ksf,Turro:2024shh,Li:2024lrl,Teplitskiy:2024gpp,Zemlevskiy:2024vxt,Ciavarella:2024lsp}.

In the quantum computation of Fig.~\ref{fig:plot12mitC},
each quantum circuit used between 40 and 50 randomized compilings with 1000 shots per compiling,
and error bars represent 95\% confidence intervals.
The time step is $\Delta\tau=0.1$.
Without self-mitigation, only the first step moves toward the true result and then all subsequent steps move in the
wrong direction, heading toward pure noise.
With self-mitigation, early time steps move toward the true result and later time steps remain there, exactly as
they should when hardware errors are mitigated successfully.
Code that ran on {\scshape ibm\_brisbane} is publicly available \cite{RandyGithub}.
Each time step used about 100 seconds of time on the quantum hardware for the combined running of the physics and self-mitigation circuits with all
randomized compilings.

Given this achievement for the square-plaquette Hamiltonian, can we use a quantum computer for the triamond Hamiltonian?
There is no obstacle in principle, but the larger plaquettes mean the ansatz of Eq.~(\ref{eq:ansatz}) is insufficient.
Inclusion of terms having three or more qubits is required, and explorations of this challenge are underway \cite{AliGithub}.
On IBM's heavy-hex architecture this leads to many swap gates because of the limited connectivity,
but the computation would be more manageable on other qubit architectures.
Overall, the implementation of larger plaquettes serves as a reminder of the added cost arising from three-dimensional physics.

\section{Real-time evolution}\label{sec:realtime}

\begin{figure}
\includegraphics[width=85mm]{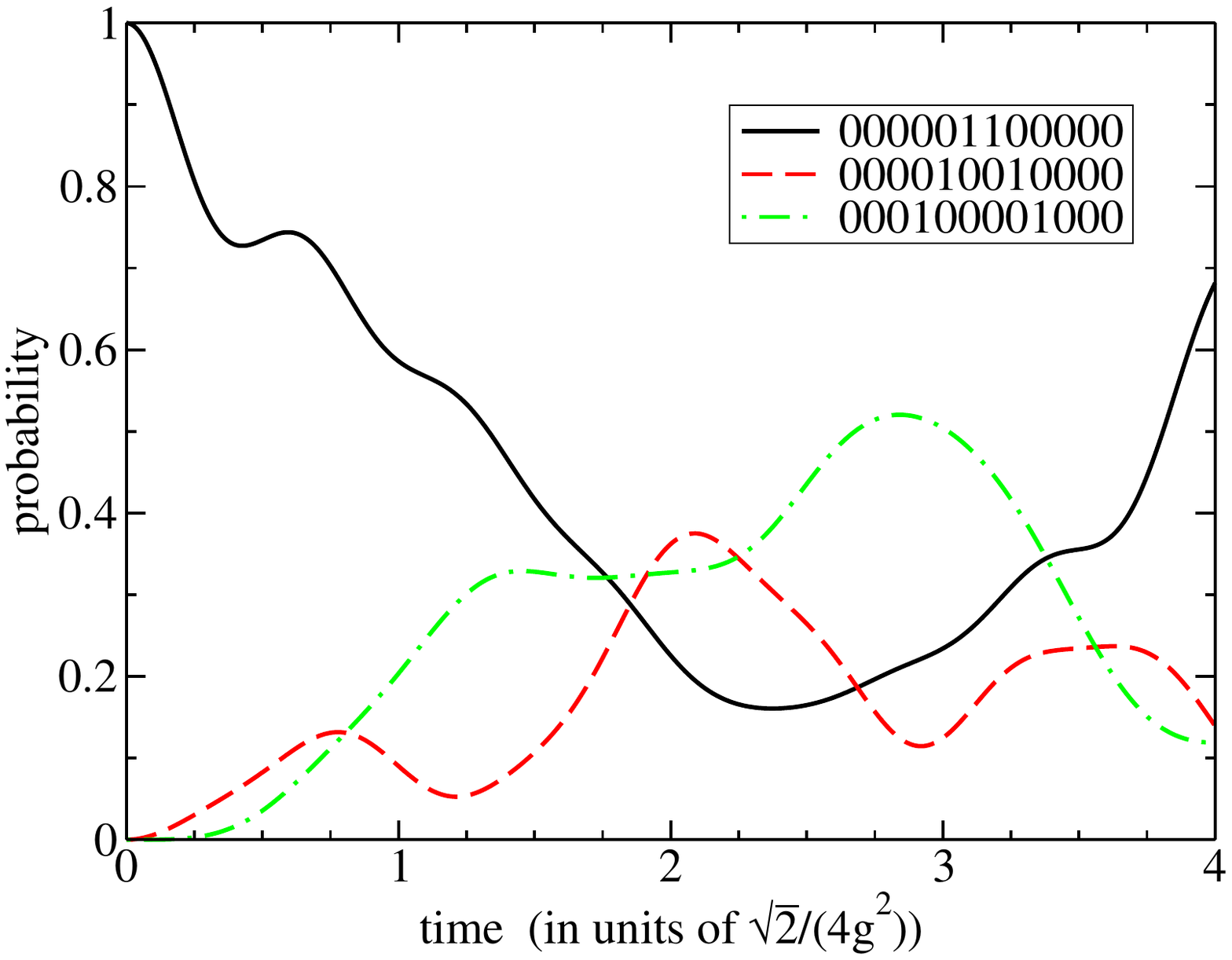}
\caption{Real-time evolution of SU(2) gauge theory on a triamond lattice of three unit cells (12 qubits)
and gauge coupling $g=1.3$.
The initial state is a bare diagonal torelon at rest.
Time evolution shows the probabilities of separation into two separate torelons, one horizontal and one vertical.
Numerical values from this graph are provided in Table~\ref{tab:truevacnumbers}.
}\label{fig:truevacuum}
\end{figure}

Consider the lowest energy state in the diagonal torelon sector of SU(2) gauge theory from Fig.~\ref{fig:spectrum}.
Because it is an energy eigenstate, its magnitude will remain unchanged during time evolution.
The dominant part of that state is simply the bare six-link torelon averaged over the 12 possible lattice locations,
\begin{eqnarray}
\left|J_0\right> &=& \frac{1}{\sqrt{12}}\bigg(\left|110000000000\right>+
\left|011000000000\right> \nonumber \\
&& +\left|001100000000\right>+
\ldots+
\left|100000000001\right>\bigg),~~~~~ \label{eq:J0}
\end{eqnarray}
which represents a bare torelon at rest.
Evolution of $\left|J_0\right>$ through real time will show that it is a superposition of several eigenstates.
The dominant eigenstates correspond to the separation of the original diagonal torelon (six gauge links)
into one horizontal and one vertical torelon (four gauge links each).
The dynamics can be visualized from Fig.~\ref{fig:LatticeLayout}(d) and numerical results are given in Fig.~\ref{fig:truevacuum}.
The calculation was done by using the exact eigenvalues and eigenvectors determined in Sec.~\ref{sec:spectrum}.
Note that all data in Fig.~\ref{fig:truevacuum} are translation invariant even through the legend shows only
one term for brevity.

According to Fig.~\ref{fig:truevacuum}, whether measurement of the qubit register is more likely to reveal a diagonal torelon or a separated pair of torelons
varies with time, as expected.
At time zero, only the diagonal torelon is present.
At a slightly later time, the closely separated pair becomes nonzero.
Shortly after that, the further separated pair becomes more probable.
This matches the classical intuition of two decay products emanating from the original single object.

Consider now the possibility of shrinking the lattice of Fig.~\ref{fig:LatticeLayout} from its length of three unit cells
down to a length of 2.75 unit cells.
Specifically, keep the first 11 square paths of Fig.~\ref{fig:LatticeLayout}(b) and erase the 12th one.
The periodic boundary condition on the flattened map now represents a twisted boundary condition for the triamond lattice.
For example, a horizontal torelon that travels completely around the long $\hat j$ direction of the lattice will return to
its original position as a vertical torelon rather than remaining horizontal.

\begin{figure}
\includegraphics[width=85mm]{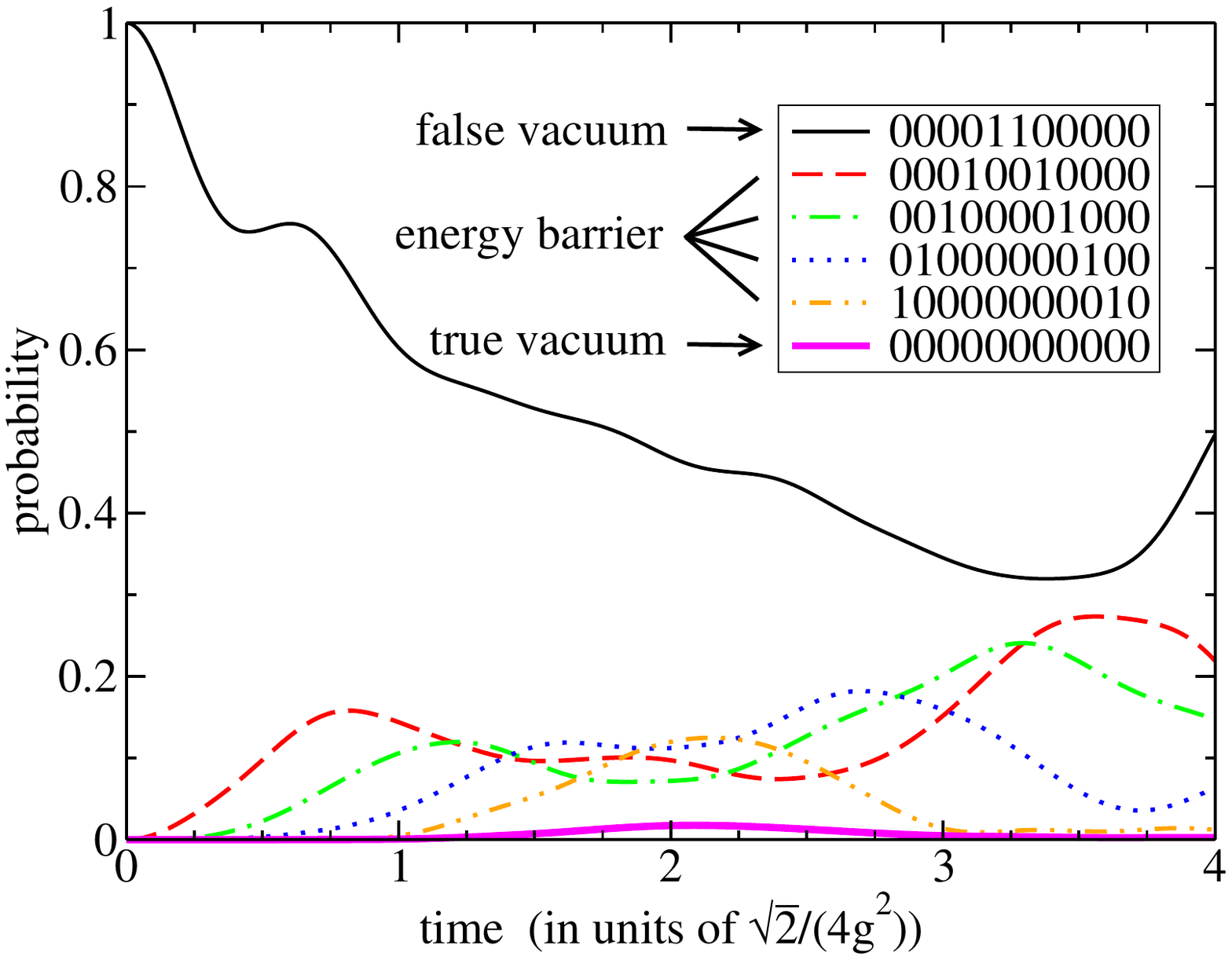}
\caption{False vacuum decay on a triamond lattice.
The lattice has $2.75$ unit cells (11 qubits) and an SU(2) gauge coupling $g=1.3$.
The initial state is a bare diagonal torelon at rest.
Real-time evolution shows the probabilities of separation into two separate torelons, and it also
shows the probability of decay into the true vacuum state with no torelons remaining.
Numerical values from this graph are provided in Table~\ref{tab:falsevacnumbers}.
}\label{fig:falsevacuum}
\end{figure}

This 11-qubit triamond lattice is an ideal setup for studying false vacuum decay.
The initial state is once again $\left|J_0\right>$ of Eq.~(\ref{eq:J0}) but with 11 terms instead of 12.
Real-time evolution will once again produce a probability for the original diagonal torelon to separate into a pair of torelons.
That pair of torelons will move further apart until they meet again after having traversed the entire $\hat j$ direction.
The new ingredient is that they can now annihilate each other because they are both horizontal or both vertical due to
the twisted boundary condition.
A direct calculation from exact eigenvalues and eigenvectors confirms these expectations as seen in Fig.~\ref{fig:falsevacuum}.
Each entry in the legend of Fig.~\ref{fig:falsevacuum} shows only one term for brevity, but all states are fully translation invariant.

The initial state of one diagonal torelon with zero momentum is our false vacuum.
It would be completely stable except for the lattice's twisted boundary condition.
The true vacuum is the state containing no torelons at all.
To make the transition from the false vacuum to the true vacuum, the original torelon needs to separate into a pair (which is
a higher energy state) and the two parts must traverse the entire lattice by travelling in opposite directions.
The growing distance between the two parts is clear from the ordering of the curves in the lower left corner of Fig.~\ref{fig:falsevacuum}.
The significant energy barrier between the false vacuum and the true vacuum results in a transition probability of about 2\%
according to Fig.~\ref{fig:falsevacuum}.

\section{Summary and outlook}\label{sec:outlook}

A gauge invariant string that winds all the way around a periodic lattice boundary is called a torelon.
Torelons are prominent objects on small lattices.
They can travel across the lattice and collide with each other due to the non-Abelian gauge interactions.
Such real-time physics is not available from traditional Euclidean lattice computations but will become readily accessible
on quantum computers by using Hamiltonian methods.

A lattice of three spatial dimensions with periodic boundary conditions has $2^3$ orthogonal sectors,
corresponding to the presence or absence of a torelon in each Cartesian direction.
Each of these independent sectors has its own stable ground state.
Applying a $\pi/2$ spatial twist at one lattice boundary is sufficient to transform a formerly stable ground state into a false
vacuum that is separated from the true vacuum by an energy barrier.

The triamond lattice is an especially efficient lattice for three spatial dimensions.
Note, for example, that the triamond lattice properly maintains the $\hat i$, $\hat j$ and $\hat k$ torelons even though
the lattice can be mapped onto the periodic ladder of Fig.~\ref{fig:LatticeLayout}(b).
In this work, we have used a small triamond lattice to calculate the spectrum of torelon states and false vacuum decay.
Also, state preparation was implemented on the quantum computer {\scshape ibm\_brisbane} for a simplified Hamiltonian through
quantum imaginary time evolution.

The three-dimensional properties of a triamond lattice are manifest in its large plaquettes, leading to a Hamiltonian with
several qubits per term that challenges the abilities of present-day quantum hardware.
The successful quantum computation of one triamond unit cell in Ref.~\cite{Kavaki:2024ijd} and the 12-qubit quantum computation
in Fig.~\ref{fig:plot12mitC} of the present work represent two steps toward this goal.

Additional studies can focus on relaxing the gauge field truncation and increasing the lattice size to ultimately
compare direct observation of false vacuum decay on a quantum computer to theoretical predictions
\cite{Braden:2018tky,Ai:2019fri,Mou:2019gyl,Zenesini:2023afv,Nishimura:2023dky,Batini:2023zpi}.
There is also a vast range of other time-dependent observables, such as particle scattering in real time, that will become
accessible through this direct quantum computing approach.

\begin{acknowledgments}
We are grateful to Matthew C.~Johnson for conversations that motivated us to pursue this topic and for reading a preliminary
draft of the manuscript.
Our work is supported in part by the Natural Sciences and Engineering Research Council (NSERC) of Canada.
We also acknowledge the use of IBM Quantum services for this work.

The views expressed are those of the authors, and do not reflect the official policy or position of IBM or the IBM Quantum team.
\end{acknowledgments}

\section*{Data Availability}

The data supporting this study's findings are available within the article.

\appendix

\section{ENERGY EIGENVALUES \\ AND EIGENVECTORS}\label{app:valsvecs}

The basis states of our computations can be represented by listing the SU(2) eigenvalues of the 12 numbered gauge links
of Fig.~\ref{fig:LatticeLayout}(b) in order from left to right.
The bare vacuum state is
\begin{equation}
\left|A\right> = \left|000000000000\right>
\end{equation}
and examples of single plaquette states include
\begin{eqnarray}
\left|B_1\right> &=& \left|101000000000\right>, \\
\left|B_2\right> &=& \left|010100000000\right>, \\
\left|B_3\right> &=& \left|001010000000\right>, \\
\vdots \nonumber \\
\left|B_{12}\right> &=& \left|010000000001\right>,
\end{eqnarray}
and
\begin{eqnarray}
\left|C_1\right> &=& \left|111100000000\right>, \\
\left|C_2\right> &=& \left|011110000000\right>, \\
\left|C_3\right> &=& \left|001111000000\right>, \\
\vdots \nonumber \\
\left|C_{12}\right> &=& \left|111000000001\right>.
\end{eqnarray}
Notice that the subscript represents screw-translation and corresponds to a symmetry of the triamond lattice.
Other significant states are
\begin{eqnarray}
\left|D_1\right> &=& \left|100010000000\right>, \\
\left|E_1\right> &=& \left|100000100000\right>,
\end{eqnarray}
together with their screw-translations.
States of definite screw-momentum can be obtained from a Fourier transform,
\begin{eqnarray}
\left|B_\theta\right> &=& \frac{1}{\sqrt{12}}\sum_{j=1}^{12}e^{-ij\theta}\left|B_j\right>, \\
\left|C_\theta\right> &=& \frac{1}{\sqrt{12}}\sum_{j=1}^{12}e^{-ij\theta}\left|C_j\right>, \\
\left|D_\theta\right> &=& \frac{1}{\sqrt{12}}\sum_{j=1}^{12}e^{-ij\theta}\left|D_j\right>
\end{eqnarray}
for $\theta\in\{0,\pm\frac{\pi}{6},\pm\frac{\pi}{3},\pm\frac{\pi}{2},\pm\frac{2\pi}{3},\pm\frac{5\pi}{6},\pi\}$
and
\begin{equation}
\left|E_\theta\right> = \frac{1}{\sqrt{6}}\sum_{j=1}^6e^{-ij\theta}\left|E_j\right>
\end{equation}
for $\theta\in\{0,\pm\frac{\pi}{3},\pm\frac{2\pi}{3},\pi\}$.
Table~\ref{tab:NoTorSpec} shows the leading contributions to the eigenvectors having the lowest energies in the no-torelon sector.

\begin{table}
\caption{The smallest energy eigenvalues in the no-torelon sector with their eigenvectors, for $g=1.3$.
Only the leading contributions to each eigenvector are listed.
Entries containing $\pm$ are degenerate pairs.
The eigenvalues are displayed graphically in Fig.~\ref{fig:spectrum}.
}\label{tab:NoTorSpec}
\begin{ruledtabular}
\begin{tabular}{ll}
value & vector \\
\hline
 7.9907 & $0.22\left|A\right> - 0.76\left|B_0\right> + 0.57\left|E_0\right> + \ldots$ \\
 7.8551 & $0.23\left|B_\pi\right> - 0.57\left|E_\pi\right> + \ldots$ \\
 7.5866 & $0.79\left|B_{\pm2\pi/3}\right> + 0.56\left|E_{\pm2\pi/3}\right> + \ldots$ \\
 7.5673 & $0.15\left|B_{\pm\pi/2}\right> + 0.96\left|D_{\pm\pi/2}\right> + \ldots$ \\
 7.5224 & $0.96\left|B_{\pm\pi/2}\right> - 0.15\left|D_{\pm\pi/2}\right> + \ldots$ \\
 7.4997 & $0.78\left|B_{\pm\pi/3}\right> - 0.57\left|D_{\pm\pi/3}\right> + \ldots$ \\
 6.5693 & $0.40\left|B_{\pm2\pi/3}\right> - 0.70\left|D_{\pm2\pi/3}\right> - 0.56\left|E_{\pm2\pi/3}\right> + \ldots$ \\
 6.5359 & $0.44\left|B_{\pm\pi/3}\right> - 0.69\left|D_{\pm\pi/3}\right> - 0.54\left|E_{\pm\pi/3}\right> + \ldots$ \\
 6.5357 & $0.70\left|B_{\pm\pi/6}\right> - 0.69\left|D_{\pm\pi/6}\right> + \ldots$ \\
 6.5133 & $0.70\left|B_{\pm5\pi/6}\right> - 0.68\left|D_{\pm5\pi/6}\right> + \ldots$ \\
 5.1688 & $0.17\left|A\right> - 0.35\left|B_0\right> - 0.68\left|D_0\right> - 0.56\left|E_0\right> + \ldots$ \\
 5.1084 & $0.42\left|B_\pi\right> + 0.68\left|D_\pi\right> + 0.55\left|E_\pi\right> + \ldots$ \\
-0.8795 & $0.95\left|A\right> + 0.29\left|B_0\right> - 0.12\left|C_0\right> + 0.05\left|D_0\right> + \ldots$
\end{tabular}
\end{ruledtabular}
\end{table}

Basis states in the horizontal torelon sector have six screw-translation locations, such as
\begin{eqnarray}
\left|F_1\right> &=& \left|010000000000\right>, \\
\left|F_2\right> &=& \left|000100000000\right>, \\
\left|F_3\right> &=& \left|000001000000\right>, \\
\left|F_4\right> &=& \left|000000010000\right>, \\
\left|F_5\right> &=& \left|000000000100\right>, \\
\left|F_6\right> &=& \left|000000000001\right>,
\end{eqnarray}
and
\begin{eqnarray}
\left|G_1\right> &=& \left|111000000000\right>, \\
\left|G_2\right> &=& \left|001110000000\right>, \\
\left|G_3\right> &=& \left|000011100000\right>, \\
\left|G_4\right> &=& \left|000000111000\right>, \\
\left|G_5\right> &=& \left|000000001110\right>, \\
\left|G_6\right> &=& \left|100000000011\right>.
\end{eqnarray}
Other important basis states in the horizontal torelon sector include
\begin{eqnarray}
\left|H_1\right> &=& \frac{1}{\sqrt{2}}(\left|110010000000\right> + \left|100110000000\right>),~~~~~~ \\
\left|I_1\right> &=& \frac{1}{\sqrt{2}}(\left|110000100000\right> + \left|100001100000\right>),~~~~~~
\end{eqnarray}
along with their screw-translations.
Table~\ref{tab:HorTorSpec} shows the leading contributions to the eigenvectors having the lowest energies in the horizontal torelon sector.
\begin{table}
\caption{The smallest energy eigenvalues in the horizontal torelon sector with their eigenvectors, for $g=1.3$.
Only the leading contributions to each eigenvector are listed.
Entries containing $\pm$ are degenerate pairs.
The eigenvalues are displayed graphically in Fig.~\ref{fig:spectrum}.
}\label{tab:HorTorSpec}
\begin{ruledtabular}
\begin{tabular}{ll}
value & vector \\
\hline
7.5222 & $0.96\left|G_\pi\right>$ \\
7.4834 & $0.94\left|G_{\pm2\pi/3}\right> + \ldots$ \\
7.3943 & $0.88\left|G_{\pm\pi/3}\right> - 0.46\left|H_{\pm\pi/3}\right>$ \\
7.2297 & $0.11\left|F_0\right> - 0.69\left|G_0\right> + 0.41\left|H_0\right> + 0.36\left|I_0\right> + \ldots$ \\
4.7162 & $0.96\left|F_\pi\right> + \ldots$ \\
4.1822 & $0.97\left|F_{\pm2\pi/3}\right> + \ldots$ \\
2.8189 & $0.97\left|F_{\pm\pi/3}\right> + \ldots$ \\
1.8586 & $0.95\left|F_0\right> + \ldots$
\end{tabular}
\end{ruledtabular}
\end{table}

Basis states in the diagonal torelon sector have 12 screw-translation locations.
The lowest energy states are dominated by
\begin{eqnarray}
\left|J_1\right> &=& \left|110000000000\right>, \label{eq:J1} \\
\left|J_2\right> &=& \left|011000000000\right>, \\
\left|J_3\right> &=& \left|001100000000\right>, \\
\vdots \nonumber \\
\left|J_{12}\right> &=& \left|100000000001\right>,
\end{eqnarray}
but we can also define
\begin{eqnarray}
\left|K_1\right> &=& \left|100100000000\right>, \\
\left|L_1\right> &=& \left|100001000000\right>, \label{eq:L1}
\end{eqnarray}
along with their screw-translations.
Table~\ref{tab:DiagTorSpec} shows the leading contributions to the eigenvectors having the lowest energies in the diagonal torelon sector.
\begin{table}
\caption{The smallest energy eigenvalues in the diagonal torelon sector with their eigenvectors, for $g=1.3$.
Only the leading contributions to each eigenvector are listed.
Entries containing $\pm$ are degenerate pairs.
The eigenvalues are displayed graphically in Fig.~\ref{fig:spectrum}.
}\label{tab:DiagTorSpec}
\begin{ruledtabular}
\begin{tabular}{ll}
value & vector \\
\hline
7.8411 & $0.58\left|K_{\pm\pi/2}\right> + 0.80\left|L_{\pm\pi/2}\right> + \ldots$ \\
7.5380 & $0.79\left|K_{\pm\pi/2}\right> - 0.57\left|L_{\pm\pi/2}\right> + \ldots$ \\
7.1800 & $0.81\left|K_{\pm2\pi/3}\right> - 0.51\left|L_{\pm2\pi/3}\right> + \ldots$ \\
7.1496 & $0.06\left|J_{\pm\pi/6}\right> + 0.79\left|K_{\pm\pi/6}\right> + 0.51\left|L_{\pm\pi/6}\right> + \ldots$ \\
7.1183 & $0.79\left|K_\pi\right> + 0.50\left|L_\pi\right> + \ldots$ \\
6.6515 & $0.48\left|K_{\pm\pi/3}\right> + 0.84\left|L_{\pm\pi/3}\right> + \ldots$ \\
5.9206 & $0.63\left|J_{\pm5\pi/6}\right> - 0.31\left|K_{\pm5\pi/6}\right> - 0.67\left|L_{\pm5\pi/6}\right> + \ldots$ \\
5.8511 & $0.97\left|J_{\pm\pi/2}\right> + \ldots$ \\
5.8075 & $0.96\left|J_{\pm2\pi/3}\right> + \ldots$ \\
5.6195 & $0.51\left|J_0\right> + 0.37\left|K_0\right> + 0.75\left|L_0\right> + \ldots$ \\
5.4622 & $0.95\left|J_{\pm\pi/3}\right> + \ldots$ \\
5.3389 & $0.92\left|J_\pi\right> + \ldots$ \\
5.3357 & $0.72\left|J_{\pm5\pi/6}\right> + 0.43\left|K_{\pm5\pi/6}\right> - 0.49\left|L_{\pm5\pi/6}\right> + \ldots$ \\
4.8470 & $0.93\left|J_{\pm\pi/6}\right> - 0.27\left|K_{\pm\pi/6}\right> + \ldots$ \\
4.3708 & $0.81\left|J_0\right> - 0.38\left|K_0\right> - 0.35\left|L_0\right> + \ldots$
\end{tabular}
\end{ruledtabular}
\end{table}

Evolution through real time can be calculated from the full set of eigenvalues and eigenvectors.
For 3 unit cells, results are shown numerically in Table~\ref{tab:truevacnumbers} and graphically in Fig.~\ref{fig:truevacuum}.
For 2.75 unit cells, results are shown numerically in Table~\ref{tab:falsevacnumbers} and graphically in Fig.~\ref{fig:falsevacuum}.
\begin{table}
\caption{Time evolution of a diagonal torelon at rest on 3 unit cells.
Labels $J$, $K$ and $L$ correspond to Eqs.~(\ref{eq:J1}-\ref{eq:L1}).
Results from this table are displayed graphically in Fig.~\ref{fig:truevacuum}.
}\label{tab:truevacnumbers}
\begin{ruledtabular}
\begin{tabular}{cccc}
time & \multicolumn{3}{c}{probability} \\
     & $J$ & $K$ & $L$ \\
\hline
0.25 & 0.8067 & 0.0304 & 0.0028 \\
0.50 & 0.7343 & 0.0824 & 0.0356 \\
0.75 & 0.7025 & 0.1309 & 0.1117 \\
1.00 & 0.5860 & 0.0902 & 0.2035 \\
1.25 & 0.5334 & 0.0540 & 0.3011 \\
1.50 & 0.4169 & 0.1070 & 0.3281 \\
1.75 & 0.3271 & 0.2225 & 0.3208 \\
2.00 & 0.2256 & 0.3620 & 0.3273 \\
2.25 & 0.1658 & 0.3443 & 0.3468 \\
2.50 & 0.1649 & 0.2645 & 0.4356 \\
2.75 & 0.1965 & 0.1569 & 0.5142 \\
3.00 & 0.2341 & 0.1246 & 0.5041 \\
3.25 & 0.3090 & 0.2103 & 0.4171 \\
3.50 & 0.3555 & 0.2344 & 0.2720 \\
3.75 & 0.4451 & 0.2291 & 0.1511 \\
4.00 & 0.6814 & 0.1395 & 0.1187
\end{tabular}
\end{ruledtabular}
\end{table}

\begin{table}
\caption{Time evolution of a diagonal torelon at rest on 2.75 unit cells.
Column headings 10, 8, 6, 4, 2, 0 correspond to the steps remaining until torelon annihilation.
Results from this table are displayed graphically in Fig.~\ref{fig:falsevacuum}.
}\label{tab:falsevacnumbers}
\begin{ruledtabular}
\begin{tabular}{ccccccc}
time & \multicolumn{6}{c}{probability} \\
     & 10 & 8 & 6 & 4 & 2 & 0 \\
\hline
0.25 & 0.8276 & 0.0320 & 0.0024 & 0.0001 & 0.0000 & 0.0000 \\
0.50 & 0.7470 & 0.0976 & 0.0227 & 0.0036 & 0.0005 & 0.0000 \\
0.75 & 0.7232 & 0.1550 & 0.0650 & 0.0132 & 0.0011 & 0.0002 \\
1.00 & 0.6028 & 0.1439 & 0.1057 & 0.0356 & 0.0048 & 0.0008 \\
1.25 & 0.5564 & 0.1133 & 0.1190 & 0.0767 & 0.0267 & 0.0032 \\
1.50 & 0.5280 & 0.0965 & 0.0941 & 0.1139 & 0.0533 & 0.0073 \\
1.75 & 0.5058 & 0.0999 & 0.0717 & 0.1158 & 0.0895 & 0.0131 \\
2.00 & 0.4685 & 0.0973 & 0.0720 & 0.1122 & 0.1198 & 0.0173 \\
2.25 & 0.4494 & 0.0790 & 0.0867 & 0.1249 & 0.1226 & 0.0166 \\
2.50 & 0.4268 & 0.0765 & 0.1269 & 0.1639 & 0.0953 & 0.0128 \\
2.75 & 0.3823 & 0.0974 & 0.1642 & 0.1812 & 0.0503 & 0.0082 \\
3.00 & 0.3453 & 0.1516 & 0.2019 & 0.1593 & 0.0140 & 0.0048 \\
3.25 & 0.3222 & 0.2279 & 0.2396 & 0.1156 & 0.0108 & 0.0038 \\
3.50 & 0.3215 & 0.2720 & 0.2184 & 0.0595 & 0.0106 & 0.0028 \\
3.75 & 0.3581 & 0.2667 & 0.1751 & 0.0361 & 0.0126 & 0.0026 \\
4.00 & 0.4960 & 0.2191 & 0.1474 & 0.0634 & 0.0122 & 0.0027
\end{tabular}
\end{ruledtabular}
\end{table}

\section{COEFFICIENTS FOR THE \\ QITE ALGORITHM}\label{app:qite}

The quantum imaginary time evolution (QITE) algorithm provides best-fit values for the 60 coefficients $(a_\bullet)_j$
within $A$ of Eq.~(\ref{eq:ansatz}) by minimizing the difference between two states,
\begin{eqnarray}
\left|\Delta_0\right> &=& \left(\frac{e^{-i\tau A}-1}{\tau}\right)\left|\psi\right>, \\
\left|\Delta\right> &=& -iA\left|\psi\right>.
\end{eqnarray}
Writing the coefficients with the simpler notation $a_1$, $a_2$, $a_3$, \ldots, $a_{60}$, the quantity to be
minimized is
\begin{equation}
\left<\Delta_0-\Delta|\Delta_0-\Delta\right> \equiv \left<\Delta_0|\Delta_0\right> + a_jb_j + a_jS_{jk}a_k
\end{equation}
which provides an expression for the coefficients
\begin{equation}
a = -(S+S^T)^{-1}b
\end{equation}
in terms of the 60-component vector $b$ and the $60\times60$ matrix $S$, both of which are real-valued.
The vector elements are
\begin{eqnarray}
(b_y)_j &=& -i\left<\psi\right|\left[\hat H,Y_j\right]\left|\psi\right> + O(\Delta\tau), \\
(b_{yx})_j &=& -i\left<\psi\right|\left[\hat H,Y_jX_{j-1}\right]\left|\psi\right> + O(\Delta\tau), \\
(b_{xy})_j &=& -i\left<\psi\right|\left[\hat H,X_{j+1}Y_j\right]\left|\psi\right> + O(\Delta\tau), \\
(b_{yz})_j &=& -i\left<\psi\right|\left[\hat H,Y_jZ_{j-1}\right]\left|\psi\right> + O(\Delta\tau), \\
(b_{zy})_j &=& -i\left<\psi\right|\left[\hat H,Z_{j+1}Y_j\right]\left|\psi\right> + O(\Delta\tau),
\end{eqnarray}
and the matrix $S$ is defined by
\begin{equation}
S_{jk} = \left<\psi\right| V_j V_k^T \left|\psi\right>,
\end{equation}
with
\begin{equation}
V_j = \left(\begin{array}{c} Y_j \\ Y_jX_{j-1} \\ X_{j+1}Y_j \\ Y_jZ_{j-1} \\ Z_{j+1}Y_j \end{array}\right).
\end{equation}
Instead of computing all 3600 matrix elements of $S$, our ansatz retains only single-qubit terms and nearest-neighbor
two-qubit terms.
In this case, all entries needed for $S+S^T$, $b$ and every term in $H_\square$ (including its three-qubit term)
can be obtained from five measurements on the quantum computer.

For the first measurement, prepare $\left|\psi\right>$ and then measure each qubit to obtain
\begin{eqnarray}
\left<\psi\right|Z_j\left|\psi\right> &=& 1 - 2P_j, \\
\left<\psi\right|Z_{j+1}Z_j\left|\psi\right> &=& 1 - 2P_{(j+1)\oplus j},
\end{eqnarray}
where $P_j$ is the probability of measuring 1 rather than 0 for the $j$th qubit,
and $P_{k\oplus j}$ is the probability that either the $j$th or $k$th qubit (not both) is 1 rather than 0.

For the second measurement, prepare $\left(\prod_{j=\text{even}}RY_j(-\frac{\pi}{2})\right)\left|\psi\right>$ and then measure each qubit to obtain
\begin{eqnarray}
\left<\psi\right|X_j\left|\psi\right> &=& 1 - 2P_j, \\
\left<\psi\right|Z_{j+1}\left|\psi\right> &=& 1 - 2P_{j+1}, \\
\left<\psi\right|Z_{j+1}X_j\left|\psi\right> &=& 1 - 2P_{(j+1)\oplus j}, \\
\left<\psi\right|Z_{j+1}X_jZ_{j-1}\left|\psi\right> &=& 1 - 2P_{(j+1)\oplus j\oplus(j-1)},
\end{eqnarray}
where $P_{l\oplus k\oplus j}$ is the probability that an odd number of the three qubits is 1 rather than 0.

For the third measurement, prepare $\left(\prod_{j=\text{even}}RY_{j+1}(-\frac{\pi}{2})\right)\left|\psi\right>$ and then measure each qubit to obtain
\begin{eqnarray}
\left<\psi\right|Z_j\left|\psi\right> &=& 1 - 2P_j, \\
\left<\psi\right|X_{j+1}\left|\psi\right> &=& 1 - 2P_{j+1}, \\
\left<\psi\right|X_{j+1}Z_j\left|\psi\right> &=& 1 - 2P_{(j+1)\oplus j}, \\
\left<\psi\right|Z_{j+2}X_{j+1}Z_j\left|\psi\right> &=& 1 - 2P_{(j+2)\oplus{j+1}\oplus j}.
\end{eqnarray}

For the fourth measurement, prepare $\left(\prod_jRY_j(-\frac{\pi}{2})\right)\left|\psi\right>$ and then measure each qubit to obtain
\begin{eqnarray}
\left<\psi\right|X_j\left|\psi\right> &=& 1 - 2P_j, \\
\left<\psi\right|X_{j+1}\left|\psi\right> &=& 1 - 2P_{j+1}, \\
\left<\psi\right|X_{j+1}X_j\left|\psi\right> &=& 1 - 2P_{(j+1)\oplus j}.
\end{eqnarray}

For the fifth measurement, prepare $\left(\prod_jRX_j(\frac{\pi}{2})\right)\left|\psi\right>$ and then measure each qubit to obtain
\begin{eqnarray}
\left<\psi\right|Y_j\left|\psi\right> &=& 1 - 2P_j, \\
\left<\psi\right|Y_{j+1}\left|\psi\right> &=& 1 - 2P_{j+1}, \\
\left<\psi\right|Y_{j+1}Y_j\left|\psi\right> &=& 1 - 2P_{(j+1)\oplus j}.
\end{eqnarray}
The computations in this appendix have been implemented successfully on {\scshape ibm\_brisbane} as discussed in
Sec.~\ref{sec:qite}.
Numerical results are provided in Fig.~\ref{fig:plot12mitC} and Table~\ref{tab:brisbane}.
Our quantum computer code for the QITE algorithm is publically available \cite{RandyGithub}.
\begin{table}
\caption{Imaginary time evolution on 12 qubits of {\scshape ibm\_brisbane}.
The Hamiltonian is Eq.~(\ref{eq:Hsquare}) with gauge coupling $g=1$ on a periodic 12-plaquette ladder.
The true ground state is obtained successfully through self-mitigation.
The eigenvalues are displayed graphically in Fig.~\ref{fig:plot12mitC}.
}\label{tab:brisbane}
\begin{ruledtabular}
\begin{tabular}{ccc}
$\tau$ & \multicolumn{2}{c}{$\left<\psi\right|H_\square\left|\psi\right>$} \\
       & without self-mitigation & with self-mitigation \\
\hline
0.1 & -3.15$\pm$0.19 & -6.51$\pm$0.20 \\
0.2 & -2.00$\pm$0.19 & -8.46$\pm$0.28 \\
0.3 & -0.65$\pm$0.20 & -8.92$\pm$0.35 \\
0.4 &  1.09$\pm$0.23 & -9.65$\pm$0.37 \\
0.5 &  2.79$\pm$0.22 & -9.08$\pm$0.50 \\
0.6 &  4.20$\pm$0.21 & -9.38$\pm$0.58 \\
0.7 &  5.15$\pm$0.15 & -9.57$\pm$0.47 \\
0.8 &  7.32$\pm$0.17 & -8.59$\pm$0.62 \\
0.9 &  7.85$\pm$0.19 & -9.22$\pm$0.85
\end{tabular}
\end{ruledtabular}
\end{table}

\end{document}